\documentstyle[aps,psfig,preprint]{revtex}
\title{Early Gedanken Experiments of Quantum Mechanics  Revisited}
\author{Yu Shi\cite{email}}
\address{Cavendish Laboratory, University of Cambridge,
Cambridge CB3 0HE, United Kingdom}
\begin{document}
\draft
\tightenlines
\maketitle
\begin{abstract}
The famous gedanken experiments of quantum mechanics  have
played  crucial roles  in developing the Copenhagen 
interpretation. They are studied  here
from the perspective of  standard  quantum mechanics, with 
no ontological interpretation involved.
Bohr's investigation of these gedanken experiments, based on   the
uncertainty relation with his interpretation, was the origin of the
Copenhagen interpretation and is still widely adopted, but is shown 
to be  not   consistent with the   quantum mechanical view.
We point out that  in most of these  gedanken experiments,  
entanglement plays a crucial role, while its buildup 
does not change the uncertainty of the concerned 
quantity in the way thought  by Bohr. 
Especially, in the gamma ray microscope and
recoiling double-slit gedanken experiments,
we expose the entanglement  based on momentum exchange. It is shown that
even in such cases,  the loss of interference
is  only  due to the entanglement with other degrees of freedom,
while the uncertainty relation argument, which  has
not  been  questioned up to now, is not right.

\end{abstract}

\section*{I. INTRODUCTION}
From 1927 to 1935, shortly after the  construction  of  
the basic formalism of  quantum mechanics, there appeared several ingenious 
gedanken experiments, culminating in Einstein, Podolsky and
Rosen (EPR) experiment \cite{epr,wheeler}, which has since 
been  a focus of  attention in researches on foundations of quantum 
mechanics, with  entanglement \cite{shrodinger}  appreciated 
as the key notion.  
As we shall  show here, entanglement also plays a crucial role in the  
other famous  gedanken experiments,  such as Heisenberg's  gamma ray
microscope \cite{heisenberg,bohr1}, Einstein's   
single-particle diffraction and 
recoiling double-slit  \cite{bohr2,jammer}, and Feynman's electron-light
scattering scheme for double-slit \cite{feynman}. 
To present, for the first time, a systematic analysis on these
famous gedanken experiments from  fully quantum mechanical point of view is
one of the purposes of this paper.

Seventy years' practice tells us to  distinguish
quantum mechanics itself, as a self-consistent  mathematical  
structure, from any  ontological interpretation.  This does not
include the sole probability interpretation,  which 
is directly connected  to observation, for instance, the probability 
distribution is given by the frequencies
of different results of an ensemble of identically prepared systems.
Whether there should be  more to say about 
an underlying ontology, or what it is, is another issue.
Furthermore, to recognize  any ontological
element  that was mixed  with quantum mechanics itself for historic 
reasons  is the first step in looking  for the right  one. Unfortunately, 
the Copenhagen school's ontological version of the uncertainty relation (UR)
is  still adopted  in many contemporary books and by many contemporary
physicists,
and Bohr's analyses of the famous gedanken experiments,  which are based on his
interpretation of the UR,  are still  widely 
accepted.  We shall comment on these analyses one by one. It turns out that
both his interpretation of the  uncertainty relation and his analyses of 
gedanken
experiments  based on the former are not  consistent with what quantum 
mechanics  says. It should be noted that the Copenhagen interpretation (CI)
 of
quantum mechanics  just originated in those analyses of gedanken experiments.
Thus  the second purpose  of this paper is to point out that CI originated
in a misconception of physics. 

The third purpose of this paper lies in the current research frontiers. 
Recently there were studies on the so-called which-way 
experiments \cite{scully,durr},
which resemble the recoiling double-slit gedanken  experiment in many
aspects. In these experiments, the interference loss is clearly due to 
entanglement, rather than uncertainty relation, because no momentum
exchange is involved. On the other hand, these results are regarded 
as another way of enforcing the so-called ``complementarity principle'',
coexisting with   Bohr's uncertainty relation arguments for
the original gedanken experiments \cite{scully,durr}. 
Here we expose the entanglement based on the momentum exchange and
show that in such a case, which does not seem to have been investigated 
in laboratories,  the interference loss is also only due to
entanglement, while the uncertainty relation arguments are wrong.
Interestingly, it will be seen that in the recoiling double-slit experiment,
the interference is not entirely lost in  Einstein's proposal of
measuring which-slit the particle passes. 

\section*{II. Uncertainty relation and photon box experiment}

We have now learned that the momentum-position  UR,
\begin{equation}
\Delta x \Delta p \geq \hbar/2, \label{ur1}
\end{equation}
  is  an  example of the general relation \cite{robertson}
\begin{equation}
\Delta A \Delta B \geq |\langle [A,B] \rangle|/2, \label{ur}
\end{equation}
 where 
$A$ and $B$ are two operators, $[A,B]=AB-BA$, 
$\Delta A=(\langle A^2\rangle-\langle A \rangle ^2)^{1/2}$,
$\langle \dots \rangle$ represents  quantum mechanical average,
which is  over a
same quantum state for different terms  in   an  inequality.
This is  only  a relation between standard deviations of the 
respective measurements following a same preparation procedure \cite{peres}:
if the same  
preparation procedure is made many times,
  but followed  by  measurements of $A$, {\em or} by
 measurements of $B$, {\em or} by  measurements of 
$[A,B]$ (if it is not a c-number),
and  the results  for $A$ have standard deviation 
$\Delta A$ while the results  for $B$ have standard deviation 
$\Delta B$, and the results for $[A,B]$ give an average 
  $\langle [A,B] \rangle$, then  $\Delta A$, $\Delta B$ and  
$\langle [A,B] \rangle$ satisfy relation (\ref{ur}).
The UR   is only a consequence of the  formalism of
quantum mechanics, rather than a basic postulate or principle.
It   has nothing to do with the accuracy 
of the  measuring instrument, nor  with the disturbance between incompatible
measurements, which are performed on different, though identically prepared,
systems.

The mathematical 
derivation  in Heisenberg's  original  paper, though 
only for a special wavefunction, is  consistent 
with the correct  meaning. But
 an ontological interpretation was given and was more emphasized on 
\cite{heisenberg,bohr1}.
In Copenhagen school's  version\cite{heisenberg,bohr1,bohr2}, 
largely justified through  the analyses of
 the  famous gedanken experiments,
it is vague
whether an  UR is an equality or 
 an inequality, 
and an  uncertainty is  interpreted 
as the objective absence or the absence of exact 
knowledge  of the value of the physical quantity, 
{\em caused by measurement or ``uncontrollable''  interaction with another
 object,
 at a given moment}. For instance, in a diffraction, the momentum uncertainty
of the particle is thought to be ``inseparately connected with the 
possibility of an exchange of momentum between the particle and the diaphragm''
\cite{bohr3}. The 
 UR  was  understood as that
{\em  determination}  of the precise 
value of 
$x$ {\em causes} a disturbance which 
destroys the precise value of $p$, and vice versa, {\em in a single 
experiment}. 

The notion of uncertainty at a given moment or in a single run of the
 experiment is beyond the formalism of quantum mechanics.
 Experimentally, 
an uncertainty  or standard deviation can only be attached
to an ensemble of experiments. Furthermore, {\em the uncertainty is determined
by the quantum state, and may remain unchanged after the interaction with
another object. In fact, in  an ideal 
 measurement, the buildup of entanglement  
 does not change  the uncertainty. Hence the uncertainty is not 
caused by an interaction with a measuring  agency.}

Now people  understand  that there does not  exist  an ``energy-time 
UR'' in the sense of (\ref{ur}) \cite{peres}.
We make  some new comments in the following.
 In Bohr's derivation \cite{bohr1,bohr2},
 one  obtains  an approximate equality  through the  
relation between  the  frequency width and the time  duration of a 
 classical wave pulse, 
or   transforms 
 the momentum-position UR to an  ``energy-time UR''
by equating an  energy uncertainty with the product of 
the momentum uncertainty and velocity, and incorrectly 
identifying time interval  or ``uncertainty'' 
as  the ratio between the 
position uncertainty  and the velocity. 
Sometimes it was a time interval, while sometimes it was a ``time
uncertainty'', that appeared in the ``energy-time UR'' \cite{bohr2}.
Therefore,  they actually provide only a  dimensional relation.
A later justification due to  Landau and Peierls 
  was from   the transition probability 
induced by a  perturbation \cite{landau1,landau2}.
For a system consisting of two parts, between which 
 the interaction is treated as a time-independent perturbation, 
the probability of a transition of the system after time $t$,
 from a state in which  energies of the two parts
 being  $E$ and $\epsilon$
to one
with energies  $E'$ and $\epsilon'$,  is proportional to
$\sin^{2} [(E'+\epsilon'-E-\epsilon)t/2\hbar]/(E'+\epsilon'-E-\epsilon)^2$.
 Thus the most probable value of $E'+\epsilon'$
satisfies 
$|E+\epsilon-E'-\epsilon'|t\sim \hbar$.
 This  interpretation   has
changed the meaning of ``energy-time UR'', and has 
 nothing to do with the general relation Eq. (\ref{ur}). Furthermore, this 
relation  only applies to the specific situation it concerns, and  
it is inappropriate  to regard  it 
  as a general relation for measurement, with
 the two parts interpreted as the
measured system and the measuring instrument. 
A counter-example was given by  Aharonov and Bohm \cite{ab}.
In fact, the relation obtained in this way must be  an equality 
with  $\hbar$ and cannot be
an inequality,
since   the   transition probability oscillates rapidly
 with
$|E+\epsilon-E'-\epsilon'|t$;
if it is $2\hbar$,
the transition probability is zero. 

In most of the gedanken experiments discussed by Bohr, the 
``energy-time UR''  was touched on 
 only as a
vague  
re-expression of the momentum-position UR.
It was directly dealt with  only in 
the  photon box experiment, proposed   by Einstein 
in sixth Solvay conference held in 1930 \cite{bohr2}.
Consider a box filled with  radiation and containing a clock 
 which controls  the opening of a shutter 
for a  time  $T$ (Fig. 1). Einstein suggested that the energy escaped 
from the hole can be measured to an arbitrary precision by weighing 
the box before and after the opening of the shutter.
In Bohr's analysis,
the box is weighed by a spring balance. He argued that for a determination of
the displacement, there is an accuracy $\Delta q$ connected to a 
momentum uncertainty $\Delta p \sim h/\Delta q$. It was thought that it should 
be smaller than the impulse given by gravitation to a mass uncertainty
$\Delta m$ during the interval $T$, i.e.  $h/\Delta q < Tg\Delta m$. 
On the other hand, $\Delta q$ is thought to be related to a  $\Delta T$
through  gravitational redshift,
$\Delta T/T=g\Delta q/c^2$, where $g$ is the gravitational constant, 
$c$ is the light velocity.
Consequently, $\Delta T\Delta E >h$, where $E=mc^2$.
One  problem in this derivation is that in the UR, the momentum uncertainty
is an intrinsic property of the box, there is no reason why it should be
smaller than $Tg\Delta m$. Another problem is that gravitation redshift 
causes a change of time, hence in the above derivation,
 $\Delta T$ corresponding to an 
uncertainty  $\Delta q$   has to  be ``an uncertainty of
the change of $T$''  if $g$ is taken as a constant. In 
contemporary established physics, both $T$ and $g$ cannot  have 
well-defined uncertainties. 
To cut the mess, 
we simply regard  Bohr's analysis   as 
no more than  giving dimensional relations.

Although we do not need to  go further, it may be noted  that the state
of the     box plus  the  inside photons is entangled  with 
the state of the  photons which have leaked out.

\section*{III.  Gamma ray microscope experiment.} 

Introduced  during  the discovery  of 
UR in 1927, the gamma ray microscope experiment
 greatly influenced the formation of CI.
The approach  of Copenhagen school 
is as follows \cite{bohr1,jammer}.
 Consider an electron  near the focus  $P$  under the
 lens (Fig. 2). 
 It  is observed through the scattering of a 
photon of  wavelength $\lambda$, thus the 
 resolving power of the microscope,  obtained from  Abbe's formula 
in classical optics {\em for a classical object},
   gives an uncertainty of any
position measurement, 
$\Delta x \sim \lambda/(2\sin \epsilon$), where $x$ is parallel to the lens,
 $2\epsilon$ is
 the angle subtended by the 
aperture of the lens at P.
  For the electron 
to be observed, the photon must be scattered into the angle $2\epsilon$, 
correspondingly  there is a latitude for the electron's momentum
after scattering, $\Delta p_{x}\sim 2h\sin\epsilon/\lambda$.
Therefore $\Delta x \Delta p_{x}\sim h$.
Heisenberg's initial  analysis, which had been acknowledged to be wrong, 
attributed the momentum uncertainty to the discontinuous momentum
change due to scattering \cite{heisenberg}.

A  quantum mechanical approach can be made  for a  general
situation. The account  is   totally different from the views of Bohr  
and Heisenberg.
 Suppose that  the state  of the electron before scattering
 is 
\begin{equation}
|\Phi\rangle_e = \int\psi (\bbox{r})|\bbox{r}\rangle_e d\bbox{r} = 
\int \phi(\bbox{p})|\bbox{p}\rangle_e d\bbox{p},
\end{equation}
and that the state of the photon   is a plane wave with a given 
momentum $\bbox{k}$,
\begin{equation}
 |\Phi\rangle_{ph} = \frac{1}{\sqrt{2\pi}}\int e^{i\bbox{k}\cdot\bbox{r}}
|\bbox{r}\rangle_{ph} d\bbox{r}=|\bbox{k}\rangle_{ph},
\end{equation}
where $\hbar$ is set to be $1$.
Before interaction, the state of the system is simply 
$|\Phi\rangle_e |\Phi\rangle_{ph}$. After interaction, 
the electron and the photon
 become entangled, until decoherence. If decoherence does not happen till 
the detection of the photon, the situation is   similar to 
  EPR experiment. If decoherence
 occurs before the photon is detected,  as in a more realistic situation,
 the observation is made on a
mixed state. 
The entangled state after scattering is
\begin{eqnarray} 
|\Psi\rangle &=& \int \int \phi(\bbox{p})C(\bbox{\delta p})
|\bbox{p}+\bbox{\delta p}\rangle_e|\bbox{k}-\bbox{\delta p}\rangle_{ph}
d\bbox{p}d(\bbox{\delta p})\\
&=& \int\int C(\bbox{\delta p})
\psi(\bbox{r}) e^{i\bbox{\delta p}\cdot\bbox{r}}
|\bbox{r}\rangle_e |\bbox{k}-\bbox{\delta p}\rangle_{ph} d\bbox{r}
  d(\bbox{\delta p}), \label{ss}
\end{eqnarray}
where $\bbox{\delta p}$ is the momentum exchange between the electron and the
photon, subject to the constraint of  energy conservation, 
 $C(\bbox{\delta p})$ represents the probability amplitude 
for each possible value of $\bbox{\delta p}$ and is  determined by the
 interaction.
Note that the states before and after scattering are  connected by 
a $S$ matrix, which depends only on the interaction Hamiltonian. 
This is because 
the interaction happens in a very local regime  
and very short time interval, i.e.
 in   the time and spatial scales  concerning us, 
we may  neglect  the time interval of the interaction. 

$|\Psi\rangle$ may be simply re-written as
\begin{equation} 
|\Psi\rangle =  \int \psi(\bbox{r}) |\bbox{r}\rangle_e 
|\bbox{r}\rangle_s  
d\bbox{r}, \label{s}
\end{equation}
where $|\bbox{r}\rangle_s=\int   C(\bbox{\delta p})
e^{i\bbox{\delta p}\cdot\bbox{r}}|\bbox{k}-\bbox{\delta p}\rangle_{ph} 
d(\bbox{\delta p})$ represents that a scattering takes  place at $\bbox{r}$.
$_{s}\langle \bbox{r'}|\bbox{r}\rangle_s$$=$$\int |C(\bbox{\delta p})|^2
e^{i\bbox{\delta p}(\bbox{r}-\bbox{r'})}d(\bbox{\delta p})$, hence 
$_{s}\langle \bbox{r}|\bbox{r}\rangle_s=1$, but $|\bbox{r}\rangle_s$
 with different $\bbox{r}$ are
 not orthogonal to each other.
One may find that {\em the position uncertainty remains unchanged while 
the momentum uncertainty becomes larger}.

More generally, if the incipient photon is also a wave-packet, i.e., 
 $|\Phi\rangle_{ph} =\int \phi_{ph}(\bbox{k})|\bbox{k}\rangle_{ph} 
d\bbox{k}$$=\int \psi_{ph}(\bbox{r})|\bbox{r}\rangle_{ph} 
d\bbox{r}$, 
then
$|\Psi\rangle=\int\int\int \phi(\bbox{p})\phi_{ph}(\bbox{k})C(\bbox{\delta p})
|\bbox{p}+\bbox{\delta p}\rangle_e|\bbox{k}-\bbox{\delta p}\rangle_{ph}
d\bbox{p}d\bbox{k}d(\bbox{\delta p})$$=$$\int\int\int C(\bbox{\delta p})
e^{i\bbox{\delta p}(\bbox{r_1}-\bbox{r_2})}\psi(\bbox{r_1})
\psi_{ph}(\bbox{r_2})|\bbox{r_1}\rangle|\bbox{r_2}\rangle_{ph}
d\bbox{r_1}d\bbox{r_2}d(\bbox{\delta p})$. 

It is clear that an uncertainty    is an intrinsic 
property determined by the quantum state, its  meaning is totally 
different from
that considered 
by Bohr from the perspective of the optical properties of the microscope such
as the resolution power as given by   classical optics. 
How  an uncertainty changes depends on the states before and
after the interaction, and it
may remain unchanged.

In an ideal  measurement  as discussed by von Neumann 
\cite{von,jammer},
 if the 
system's state is $\sum_k \psi_k\sigma_k$, then after interaction with the 
apparatus, the system-plus-apparatus has the entangled state
$\sum_k \psi_k\sigma_k\alpha_k$. $\sigma_k$ and $\alpha_k$ are orthonormal sets
of the system and apparatus, respectively. In such a case, the expectation 
value of any observable of the system and thus its  uncertainty
is  not changed by the interaction with the apparatus.

\section*{IV.  Detection of a  diffracted particle}

At the Fifth Solvay Conference held in 1927,
Einstein considered the diffraction of  
a  particle  from a slit  to a hemispheric screen \cite{jammer} (Fig. 3).
He  declared that  if the wavefunction represents  an ensemble of
particles distributed in space rather than one particle, then
$|\psi(\bbox{r})|^2$ expresses the percentage of particles presenting
at $\bbox{r}$.  But  if quantum mechanics describes individual processes,
$|\psi(\bbox{r})|^2$ represents the probability that at  a given moment a
same particle shows its presence at $\bbox{r}$.  Then as long as no 
localization is effected, the particle has the possibility over the whole
area of the screen; but as soon as it is localized, a 
peculiar instantaneous  action at a
distance must be assumed to take place which prevents 
the continuously distributed wave from producing an effect at two places 
on the screen.

By definition,
the concept of probability implies the concept of ensemble, 
which means the  repeat of identically prepared processes.
Therefore as far 
as $\psi(\bbox{r})$ represents a probability wave rather than a physical
wave in the real space,  it makes  no difference  whether 
 it  is understood 
in terms of an ensemble or in terms of a single particle. 
What Einstein referred to 
 here by ensemble is effectively the classical ensemble,
i.e. only the probability  plays a role while $\psi(\bbox{r})$ 
does not matter directly.
The essential 
problem of Einstein is
 how a classical event originates from the
quantum state, which was unsolved  in the early years and 
remains, of course,
an important and active subject. 
In the fully quantum mechanical view, 
 the diffraction is not essential 
for Einstein's problem. 

Here comes the entanglement between the detector and the particle.
 The state of the combined  system evolutes from the product of those of
the particle and the screen into an entangled state 
\begin{equation}
|\Psi\rangle = \int \psi(\bbox{r}) |\bbox{r}\rangle_p|\bbox{r}\rangle_d 
d\bbox{r},
\end{equation}
where $|\bbox{r}\rangle_p$ is the position eigenstate of the particle,
$|\bbox{r}\rangle_d$ represents that a particle is detected at
$\bbox{r}$.
The measurement result of  the particle position
is described by a classical ensemble, with the 
diagonal density matrix
\begin{equation}
\rho_p = \int |\psi(\bbox{r})|^2|\bbox{r}\rangle_{pp}\langle\bbox{r}|d\bbox{r}.
\label{pd}
\end{equation}

von Neumann formulated this as a  postulate
 that in addition to the unitary evolution,
 there is a 
nonunitary, discontinuous  
``process of the first kind'' \cite{von},
which cancels the off-diagonal terms in   the   pure-state density matrix
$|\Psi\rangle\langle \Psi|$,
 leaving
a diagonal reduced density matrix
$\rho_r$$=$$\int |\psi(\bbox{r})|^2 |\bbox{r}\rangle_{pp}\langle \bbox{r}|
 |\bbox{r}\rangle_{dd}\langle \bbox{r}|d\bbox{r}$,
which implies  $\rho_p$. Equivalently the projection 
postulate may also apply directly 
 to the particle state
  to obtain (\ref{pd}). 
There are various alternative 
approaches to  this problem. 
Nevertheless,  the projection postulate should   effectively
valid in most situations.

What did Bohr say?
Instead of addressing the event  on  the {\em screen},
Bohr   discussed  the correlation between the 
particle and the {\em diaphragm} using  his version of 
  UR  \cite{bohr2} .
Einstein's problem was conceived  as to what extent a control of the
momentum and energy transfer can be used for a specification of the state
of the  particle after passing through the hole.
This is a misunderstanding:
even if its  position  on 
 the diaphragm is specified,  the particle is still not in a momentum 
eigenstate, moreover,  the particle 
still has  nonzero position wavefunction at 
every point  on the screen after arrival;
what is lost is the interference 
effect. 

\section*{V.  Recoiling double-slit and Feynman's electron-light scattering}

After Bohr's argument of the single-slit diffraction, Einstein proposed the
recoiling double-slit arrangement \cite{bohr2,jammer}.
Consider   identically prepared  particles
which, one by one,  are
incident on 
   a diaphragm  with two slits, and then 
arrive at a screen. Einstein argued against UR   as follows.

As shown in Fig. 4, midway between a stationary diaphragm $D_1$ with a
single slit as the particle source 
and a screen $P$, a movable diaphragm $D_2$ is suspended 
by a weak spring $Sp$. 
The two  slits $S'_2$ and $S''_2$  are separated by a distance $a$,
much smaller than $d$, the distance between $D_1$ and $D_2$. Since 
the momentum imparted to $D_2$ depends on 
 whether the particle passes 
through 
 $S'_2$ or $S''_2$, hence the  position in passing through the slits 
can be determined.
On the other hand, the momentum of the particle 
can be measured from  the interference
 pattern.

Bohr 
pointed out
that 
 the two paths'
difference of momentum transfer is   $\Delta p=p\omega$ (this is a 
fault, in this setup, it should be  $2p\omega$), where 
$\omega$  is the angle subtended by the two slits  at the single slit in 
$D_1$. He argued that any measurement of momentum with an accuracy sufficient
to decide $\Delta p$ must involve a position  uncertainty 
of at least $\Delta x=h/\Delta p=
\lambda/\omega$, which equals 
the width
of the interference fringe. Therefore the momentum determination of $D_2$ 
for the decision  of
 which path the particle
passes involves a position uncertainty which
destroys  the interference.
 This was regarded as a typical   ``complementary phenomenon''
 \cite{bohr2}.

In Feynman's light-electron scattering scheme, which slit the electron passes
is observed by the scattering with a photon. One usually
 adopts Bohr's analysis above   by replacing the momentum
of the diaphragm as that of the photon. 

{\em Bohr's  argument means  that 
in determination of which slit the particle passes, its momentum uncertainty
becomes smaller enough}. Clearly this may not happen. For instance, 
as we have seen in 
our  analysis on the gamma ray microscope, 
 scattering with a plane wave  increases the momentum
uncertainty on the contrary.
This is an indication that Bohr's analysis is not correct.

Bohr's reasoning 
is avoided in a  proposal using   two laser cavities   as the 
which-slit  tag for an atomic beam \cite{scully}, and in a recent 
 experiment  where the internal electronic 
state of the atom acts as the  which-slit  tag \cite{durr,knight}.
However,  as a showcase of the contemporary influence of 
CI, there was no doubt on  
Bohr's  argument  
in the original gedanken experiment, and the current  experimental results 
 were  framed in terms of Copenhagen ideology,  even the debate
on whether
UR plays a role in such which-way experiments was titled as
``Is complementarity more fundamental than the uncertainty principle?''
\cite{durr}.
We shall show that it is a universal mechanism that interference is 
destroyed by entanglement with another degree of freedom. Here
the momentum exchange is just the basis for the entanglement. 
Bohr's analysis is not consistent with the  fully 
quantum mechanical account. 

We make a 
general account  applicable 
to both single-slit diffraction and many-slit interference.
Let us assume that  just before diffraction by the slit(s),
the state of the particle is 
\begin{equation}
|\Phi(t_0-0)\rangle_p
= \int\psi(\bbox{r_0},t_0)|\bbox{r_0}\rangle_p d\bbox{r_0}
= \int\phi(\bbox{p},t_0)|\bbox{p}\rangle_p d\bbox{p}
\end{equation}
 where $\bbox{r}_0$ belong to  the slit(s).
After diffraction, the state  is
\begin{equation}
|\Phi(t>t_0)\rangle_p =\int \psi(\bbox{r},t) |\bbox{r}\rangle_p
d\bbox{r},\label{ad}
\end{equation}
with 
\begin{equation}
\psi(\bbox{r},t)=\int 
 \psi(\bbox{r_0},t_0)G(\bbox{r},t;\bbox{r_0},t_0)d\bbox{r_0}, 
\end{equation}
where $G(\bbox{r},t;\bbox{r_0},t_0)$ is a propagator. 
The interference  appears since 
the probability that the diffracted 
particle is at $\bbox{r}$ is 
$|\int\psi({\bbox{r}_0,t_0)G(\bbox{r},t;\bbox{r}_0,t_0})
d\bbox{r}_0|^2$, instead of 
$\int|\psi({\bbox{r}_0,t_0)G(\bbox{r},t;\bbox{r}_0,t_0})|^{2}
d\bbox{r}_0$. 
The diffraction does  not change  the uncertainties of  
position and  momentum, as seen from  
$|\Phi(t\rightarrow  t_0)\rangle\rightarrow |\Phi(t_0-0)\rangle$.

Before interacting  with the photon ($t< t_0$), the
 diaphragm has a definite position $\bbox{r_i}$. This is a gedanken 
experiment, in which the relevant degrees of freedom are well isolated.
 Hence the diaphragm is described by the quantum state
\begin{equation}
|\Psi(t< t_0)\rangle_d=|\bbox{r_i}\rangle_d=\int \delta(\bbox{r}-\bbox{r_i})
|\bbox{r}\rangle_d d\bbox{r}=
\frac{1}{\sqrt{2\pi}}
\int e^{-i\bbox{k}\cdot\bbox{r_i}}|\bbox{k}\rangle_d d\bbox{k}, \label{del}
\end{equation}
and the state of the combined 
 system of particle-plus-diaphragm is 
$|\Phi(t< t_0)\rangle_p|\bbox{r_i}\rangle_d$.
If the diaphragm is fixed, the state of whole system after diffraction 
is  the product of  $|\bbox{r_i}\rangle_d$ 
and the state  of the particle
 as given by Eq. (\ref{ad}).

If the diaphragm is moveable, 
after the interaction, the state of the combined 
system  is an entangled one. Right after the scattering,
\begin{eqnarray}
|\Psi( t_0+0)\rangle &=& \frac{1}{\sqrt{2\pi}}\int\int\int C(\bbox{\delta p})
\phi(\bbox{p},t_0)e^{-i\bbox{k}\cdot\bbox{r_i}}
|\bbox{p}+\bbox{\delta p}\rangle_p|\bbox{k}-\bbox{\delta p}\rangle_d
d(\bbox{\delta p})d\bbox{p}d\bbox{k} \\
&=& \int\int C(\bbox{\delta p})\psi(\bbox{r_0},t_0)
e^{i\bbox{\delta p}\cdot\bbox{r_0}}|\bbox{r}_0\rangle_p
e^{-i\bbox{\delta p}\cdot\bbox{r_i}}|\bbox{r_i}\rangle_d
d(\bbox{\delta p})d\bbox{r_0}.
\end{eqnarray}
Then, 
\begin{eqnarray}
|\Psi(t > t_0\rangle &=& \int\int\int C(\bbox{\delta p})
\psi(\bbox{r_0},t_0)G(\bbox{r},t;\bbox{r_0},t_0)
e^{i\bbox{\delta p}\cdot\bbox{r_0}}|\bbox{r}_0\rangle_p
U(t,t_0)e^{-i\bbox{\delta p}\cdot\bbox{r_i}}|\bbox{r_i}\rangle_d
d(\bbox{\delta p})d\bbox{r_0}d\bbox{r},\\
&=& \int \int \psi(\bbox{r_0},t_0)G(\bbox{r},t;\bbox{r_0},t_0)|\bbox{r_0}\rangle_p
|\bbox{r_0}\rangle_s(t)d\bbox{r_0}d\bbox{r}, \label{fey}
\end{eqnarray}
where $U(t,t_0)$ represents the evolution of the diaphragm state, and
\begin{equation}
|\bbox{r_0}\rangle_s(t)= \int C(\bbox{\delta p})
e^{i\bbox{\delta p}\cdot\bbox{r_0}}
U(t,t_0)e^{-i\bbox{\delta p}\cdot\bbox{r_i}}
|\bbox{r_i}\rangle_d d(\bbox{\delta p}).
\end{equation}
Generally speaking,
this  entanglement is not maximal,  hence the interfere is 
not completely destroyed. 

In Feynman's electron-light scattering scheme,  we may suppose that
the scattering 
takes  place at the slit(s).
In general, the photon is a wave packet  
 $|\Phi\rangle_{ph}=\int \phi_{ph}(\bbox{k})|\bbox{k}\rangle_{ph} d\bbox{k}$
$=$$\int \psi_{ph}(\bbox{r})|\bbox{r}\rangle_{ph} d\bbox{r}$,
 then
$|\Psi(t>t_0)\rangle$$=$$\int\int\int\int C(\bbox{\delta p})
\psi(\bbox{r_0})G(\bbox{r},t;\bbox{r_0},t_0)
e^{i\bbox{\delta p}\cdot\bbox{r_0}}|\bbox{r}_0\rangle_p
U(t,t_0)e^{-i\bbox{\delta p}\cdot\bbox{r_1}}
\psi(\bbox{r_1})|\bbox{r_1}\rangle_{ph}
d(\bbox{\delta p})d\bbox{r_0}d\bbox{r_1}d\bbox{r}$.
Again,   the interference is not  completely destroyed.
If the photon is a plane wave, as
 we have known in the above discussions on 
the gamma ray microscope, 
 the  position uncertainty remains unchanged while the
momentum uncertainty increases, contrary to Bohr's claim. 

In general, in both the moveable diaphragm and
the photon scattering schemes, the change of the uncertainty is dependent
of the states before and after the interaction. It is not right to simply
say that the momentum uncertainty becomes smaller, as thought by Bohr.  
One should also note that there are various possible
momentum exchanges  $\bbox{\delta p}$, subject to energy conservation, and this
is {\em independent} of the position
 on the diaphragm, or which slit the particle
 passes in the double-slit experiment. {\em In general, 
both  before and after the interaction with the diaphragm, the particle is in a
superposition of different momentum eigenstates}. Even after the detection of
the particle on the screen,  the states of the diaphragm
and particle 
still do not reduce to those 
 with a definite momentum exchange. This  was not appreciated by
either Bohr or Einstein, and 
 is another point  inconsistent  with 
Bohr's analysis,  which is based on a classical picture supplemented by
an uncertainty relation.
 
\section*{VI. Concluding remarks}
Regarding Copenhagen school's view of uncertainty, one  cannot directly
prove or disprove  the notion of uncertainty at a given moment in a single
run of experiment, which  is  beyond the 
standard formalism of  quantum mechanics. 
However, it is clearly wrong to attribute  the uncertainty  to the 
interaction with a ``measuring agency''. It is also wrong to regard the 
uncertainty as a bound for the accuracy of the measuring instrument, given 
by  classical physics, as done in Bohr's analyses of gamma ray microscope 
and recoiling double-slit.
On the other hand, it is inappropriate to regard 
the consequence of the interaction   simply as 
 causing  the  uncertainty,    while neglect 
the buildup of the entanglement, which may not change the uncertainty,
or may change it  but not in the way thought  by Bohr.

We have seen that Bohr's analyses of the gedanken experiments are not 
consistent with the   quantum mechanical accounts. 
This indicates that  the essence of quantum mechanics
 cannot be simply reduced to a classical picture supplemented by
an uncertainty relation.
More weirdness of quantum phenomena comes from the superposition,
especially the entanglement. 
The crucial importance of entanglement  in quantum mechanics was not
well appreciated  in the early gedanken experiments, with the attention 
focused on uncertainty relation in Copenhagen school's version.
However, it was finally  exposed 
 in EPR experiment,  and had been noted in an earlier paper
 \cite{etp}.

Had Bohr been aware of this, he   might say 
that the entanglement versus  interference implements the ``complementary 
principle'', and   might be happy that more commonness exists between 
the early gedanken experiments and EPR-experiment than  discussed 
in his reply to EPR paper\cite{bohr3}, where 
most of the  discussions deal with the ``mechanical disturbance''
in the diffraction experiment by using uncertainty relation argument; 
regarding EPR, it is only said that ``even
 at this stage there is essentially the 
question of {\it an influence on the very conditions which define the possible
 types of predictions regarding the future behavior  of the system}''.
 Note that
if the ``very condition'' refers to the quantum state, it
is just the basis of the discussions of EPR, and that  
 Einstein did  consider  it as 
possible to relinquish the assertion that 
``the states of spatially 
separated objects are independent on each other'', but with wavefunction 
regarded as a description of an ensemble \cite{ein2}.
The wording of the Copenhagen interpretation is  so vague and flexible 
that one can easily embed new meanings, as people have been doing in the
past many years. However, no matter how to refine its meaning, the
 ``complementary principle''  does not provide  any
better understanding than that provided by quantum mechanics itself. 
Afterall,
the language of physics is mathematics rather than  philosophy. 
In fact, decoherence based on entanglement
 could have been studied in the early
days,  had not there been the advent of the Copenhagen 
interpretation \cite{kiefer},
 which   originated in  
the misconception on the early gedanken experiments.   

\bigskip
I thank Claus Kiefer for useful discussions. 
\newpage
{\bf Figures}

Figure 1. Photon box. Copied from Ref. [6].

Figure 2. Gamma ray microscope. Copied from Ref. [7].

Figure 3. Detection of diffracted particle. Copied from
Ref. [7].

Figure 4. Recoiling double-slit. Copied from Ref. [7].

\psfig{figure=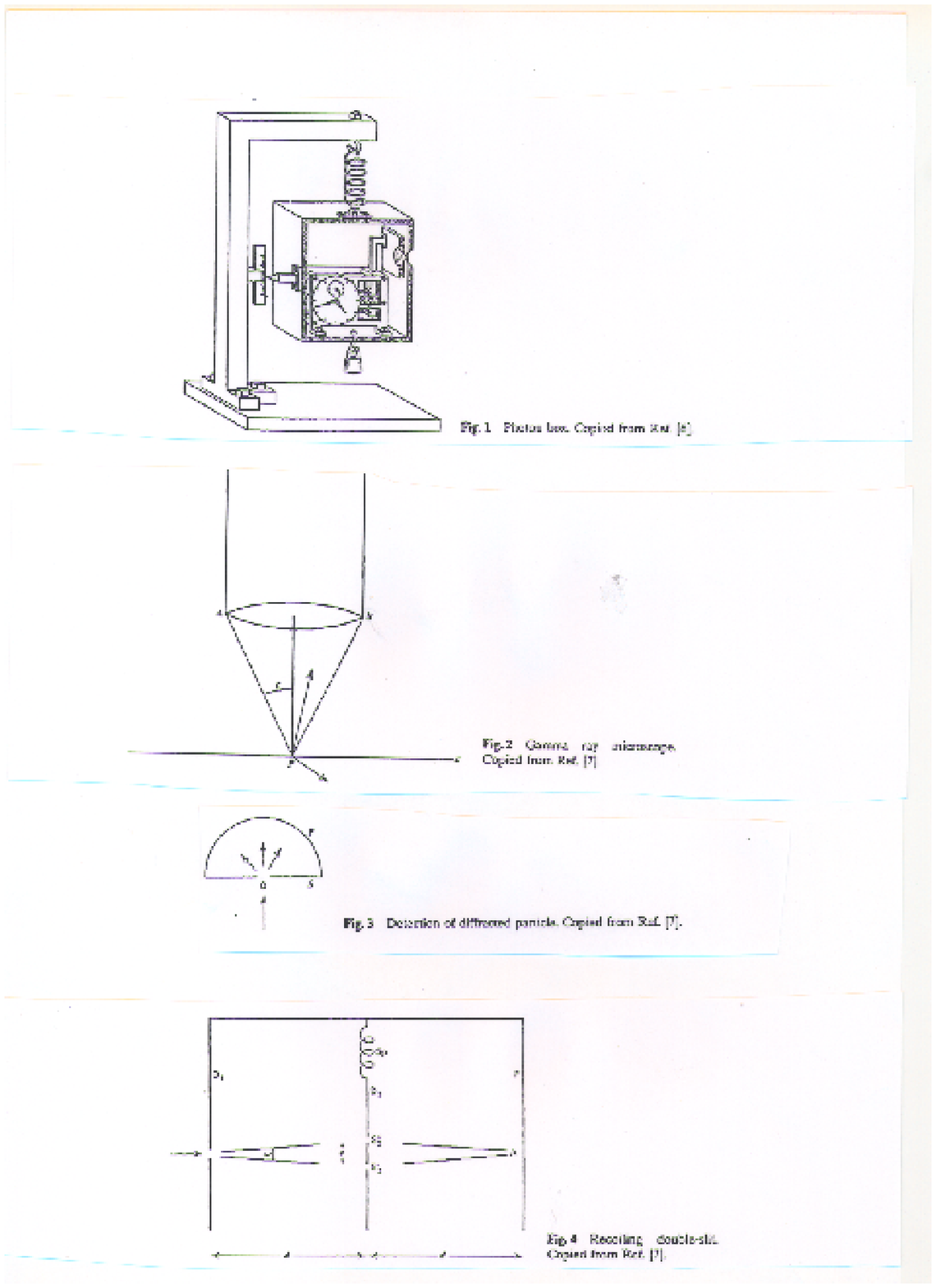}


\begin{references}
\bibitem[*]{email} Email address: ys219@phy.cam.ac.uk
\bibitem{epr} A. Einstein, B. Podolsky, and  N. Rosen,
 Phys. Rev.
{\bf 47}, 777 (1935).
\bibitem{wheeler} References 
\cite{epr,shrodinger,heisenberg,bohr1,bohr2,bohr3,robertson,landau1,etp}
 are
 collected in 
J. A.  Wheeler  and W. H. Zurek,
{\em Quantum Theory and Measurement}
 (Princeton University Press, 1983).
\bibitem{shrodinger} E. Schr\"{o}dinger,  
 Naturwissenschaften
{\bf 23}, 807; 823; 844 (1935); 
translated by J. D. Trimmer in Proc. Am. Phi. Soc. 
{\bf 124}, 323 (1980).
\bibitem{heisenberg} W. Heisenberg,  
 Z. Phys. {\bf 43}, 172  (1927); 
translated by J. A. Wheeler and W. H. Zurek
in Ref. 2.
\bibitem{bohr1} N. Bohr, 
 Nature {\bf 121}, 580 (1928). 
\bibitem{bohr2} N. Bohr, 
in {\em Albert Einstein: Philosopher-Scientist},
edited by P. A. Schilpp (Library of Living Philosophers, Evanston,
 1949).
\bibitem{jammer} M. Jammer, {\em The Philosophy of Quantum Mechanics}
(John Wiley \& Sons, New York, 1974).
\bibitem{feynman} R. Feynman, R. Leighton,  and  M. Sands, {\em The
Feynman Lectures on Physics Vol. III}
(Addison Wesley, Reading, 1965).   
\bibitem{scully} M. O. Scully,  B. G. Englert, and H. Walther,
 Nature {\bf 351}, 111 (1991).
\bibitem{durr} S. D\"{u}rr, T. Nonn, and  G. Remper,
 Nature
{\bf 395}, 33 (1998).
\bibitem{knight} P. Knight,  Nature
{\bf 395}, 12 (1998).
\bibitem{bohr3} N. Bohr,  
Phys. Rev. {\bf 48}, 696 (1935).
\bibitem{robertson} H. P. Robertson,
 Phys. Rev. {\bf 34}, 163
(1929).
\bibitem{peres} A. Peres, {\em Quantum Theory: Concepts and Methods}
(Kluwer Academic Publishers, Dordrecht, 1993).
\bibitem{landau1} L. D. Landau and 
 R. Peierls,   Z. Phys. {\bf 69}, 56 (1931);
translated in {\em Collected papers of Landau}, edited by
D. ter Haar 
(Gorden and Breach, New York, 1965),  pp. 40-51.
\bibitem{landau2} L. Landau and  E. M. Lifshitz,
{\em Quantum Mechanics},
translated by J. B. Sykes and J. S.  Bell,
(Pergamon, Oxford, 1977).
\bibitem{ab} Y. Aharonov,  and D.  Bohm,  Phys. Rev. {\bf 122},
1649 (1961).
\bibitem{von} J. von Neumann, {\em Mathematische Grundlagen der 
Quantenmechanik} (Springer-Verlag, Berlin, 1932).
\bibitem{etp} A. Einstein, R. C.  Tolman and  B. Podolsky,   Phys. Rev.
{\bf 37}, 780 (1931).
\bibitem{ein2} A. Einstein, 
 in {\em Albert Einstein: Philosopher-Scientist},
edited by  P. A. Schilpp
(Library of living philosophers, Evanston, 1949), p. 682.
\bibitem{kiefer} D. Giulini, E. Joos, C. Kiefer, J. Kupsch,
I.-O. Stamatescu and H. D. Zeh,
{\em Decoherence and The Appearance of a Classical 
World in Quantum Theory} (Springer, Berlin, 1996);
 C. Kiefer and E. Joos, in {\em Quantum Future}, 
ed. P. Blanchard and A. Jadczyk (Springer, Berlin, 1998).
\end{references}
 \end{document}